%% file: turnbench.tex
\documentclass[conference]{IEEEtran}
\IEEEoverridecommandlockouts
\usepackage{cite}
\usepackage{amsmath,amssymb,amsfonts}
\usepackage{graphicx}
\usepackage{array}
\usepackage{makecell}
\usepackage{caption}
\usepackage[hidelinks]{hyperref}
\hypersetup{pdftitle={TurnBench: A Multi-Domain Benchmark for Turn-Taking Dynamics in Spoken Dialogue}, pdfauthor={Freeman Jiang et al.}}

\newcommand{\ctype}[1]{\textit{#1}} 

\usepackage{pifont}
\newcommand{\cmark}{\ding{51}}
\newcommand{\xmark}{\ding{55}}
\usepackage{threeparttable}  
\newcommand{\evt}[1]{\textsc{#1}}

\usepackage{textcomp}
\usepackage{xcolor}
\usepackage{booktabs}
\usepackage{url}
\usepackage{xspace}
\usepackage{scrextend}
\usepackage{balance}
\newif\ifarxiv \arxivfalse
\ifarxiv
\usepackage{fancyhdr}
\fancypagestyle{arxivfirst}{\fancyhf{}%
  \fancyfoot[C]{\raisebox{0pt}[0pt][0pt]{\parbox[t]{\textwidth}{\scriptsize\raggedright
  Accepted to IEEE SLT 2026. \copyright~2026 IEEE.}}}}
\fi
\DeclareFontShape{OT1}{ptm}{m}{scit}{<->ssub * ptm/m/it}{}
\newcommand{\bench}{\textsc{TurnBench}\xspace}

\def\BibTeX{{\rm B\kern-.05em{\sc i\kern-.025em b}\kern-.08em
    T\kern-.1667em\lower.7ex\hbox{E}\kern-.125emX}}
\begin{document}
 
\title{\bench: A Multi-Domain Benchmark for Turn-Taking Dynamics in Spoken Dialogue}

\author{\IEEEauthorblockN{Freeman Jiang$^{1}$, Ramon Sanabria$^{1}$, Soham Deshmukh$^{1}$, Bandhav Veluri$^{1}$,
Simon Michael Vuch Williams$^{2}$,\\ Elliott K. Suen$^{2}$, Garreth Lee$^{2}$, Kevin Yoonho Choi$^{2}$, Takuya Umeki$^{6}$,
Riku Kubo$^{6}$, Sathvik Udupa$^{7}$,\\ Chien-yu Huang$^{3}$, Shih-Yun Shan Kuan$^{4}$, Zhuoyan Tao$^{3}$, Satyapriya Krishna$^{1}$,
Sefik Emre Eskimez$^{1}$,\\ Yu Tsao$^{5}$, Hung-yi Lee$^{4}$, Shinji Watanabe$^{3}$}
\IEEEauthorblockA{\rule{0pt}{22pt}\textit{$^{1}$Sesame AI \quad $^{2}$Mundo AI \quad $^{3}$Carnegie Mellon University \quad $^{4}$National Taiwan University}\\
\textit{$^{5}$Academia Sinica \quad $^{6}$Oto \quad $^{7}$Brno University of Technology}}
}

\maketitle
\ifarxiv\thispagestyle{arxivfirst}\fi

\begin{abstract}
Speakers in natural conversation take turns speaking and listening, deciding in real time when to take, hold, or yield the floor. However, turn-taking evaluation remains limited due to the lack of a consistent, linguistically grounded evaluation protocol and hand-annotated data covering diverse conversation types. To address this, we present \bench, a multi-domain benchmark that pairs a 30-hour, hand-labeled corpus of dyadic human conversation with a standardized evaluation protocol for end-of-turn and interruption detection. We set conversation type as a controllable experimental variable, covering six distinct interaction styles, and triple-annotate each conversation. Benchmarking 14 heterogeneous turn-taking systems, we find end-of-turn recall stable across types, while interruption false positives are strongly type-dependent and concentrated in backchannel-dense interaction styles. Although in smooth floor transfers human listeners begin speaking a median 151\,ms before the current turn ends, no current system performs equivalently without incurring excessive false positives. We release our corpus, a 104-hour training set, and a public leaderboard with an interactive dataset viewer at \textcolor{blue}{\href{https://turnbench.sesame.com}{turnbench.sesame.com}}.
\end{abstract}

 
\input{intro}
\input{relwork}
\input{corpus}

\input{eval}
\input{baselines}
\input{results}
\input{conclusion}
\input{ethics}
\section{Acknowledgment}
\looseness=-1
We thank Siddhant Arora (Meta) for early design feedback. This work used Bridges-2 at PSC and Delta and DeltaAI at NCSA through allocations CIS210014 and IRI120008P from the Advanced Cyberinfrastructure Coordination Ecosystem: Services \& Support (ACCESS) program, which is supported by National Science Foundation grants \#2138259, \#2138286, \#2138307, \#2137603, and \#2138296. Sathvik Udupa is supported by Technology Agency of the Czech Republic (TACR) project No.\ TQ28000003 ``National Center for Artificial Intelligence.'' The authors used LLMs to assist with code development. All AI-assisted content was reviewed, edited, and verified by the authors.
\newpage

\balance
\bibliographystyle{IEEEtran}
\bibliography{turnbench}
 
\end{document}

%% file: intro.tex
\section{Introduction}

In spoken dialogue, speakers continuously navigate the conversational floor through turn-taking. Interlocutors signal with linguistic and prosodic cues whether they intend to start, continue, or yield a turn, producing backchannels and managing overlaps \cite{sacks1974simplest}. Spoken dialogue research has long studied these phenomena \cite{skantze2021review} by decomposing them into discrete tasks such as end-of-turn detection \cite{udupa2026endpoint}, interruption handling, and backchannel prediction \cite{lala2017attentive, lala2019smooth}. However, these turn-taking phenomena are distinguished by overlapping cues in the same preceding speech \cite{gravano2011turn, ward2000prosodic}. Thus, these tasks are coupled and should be defined and scored against a single annotation of the conversation, instead of in isolation.

Full-duplex spoken dialogue models \cite{nguyen2023dgslm, defossez2024moshi} are now capable of negotiating the floor implicitly, but they expose no direct turn-taking decisions. There is currently no open resource for full-duplex turn-taking evaluation across conversation types. Large untranscribed corpora were built for synthesis training and have no turn-taking annotations~\cite{otospeech2025}, and the corpora that are annotated each cover only a single register \cite{reece2023candor, sheikh2025sssd, xiao2025casper, garofolo1993timit, panayotov2015librispeech, paul1992wsj, godfrey1992switchboard, cieri2004fisher, anderson1991maptask, carletta2006ami, janin2003icsi, dubois2000santabarbara, barker2018chime5, vinnikov2024notsofar, vansegbroeck2020dipco, sanabria2023edacc} (Section~\ref{sec:related}, Table~\ref{tab:corpora-comparison}).

\begin{figure}[t]
\centering
\includegraphics[width=\columnwidth]{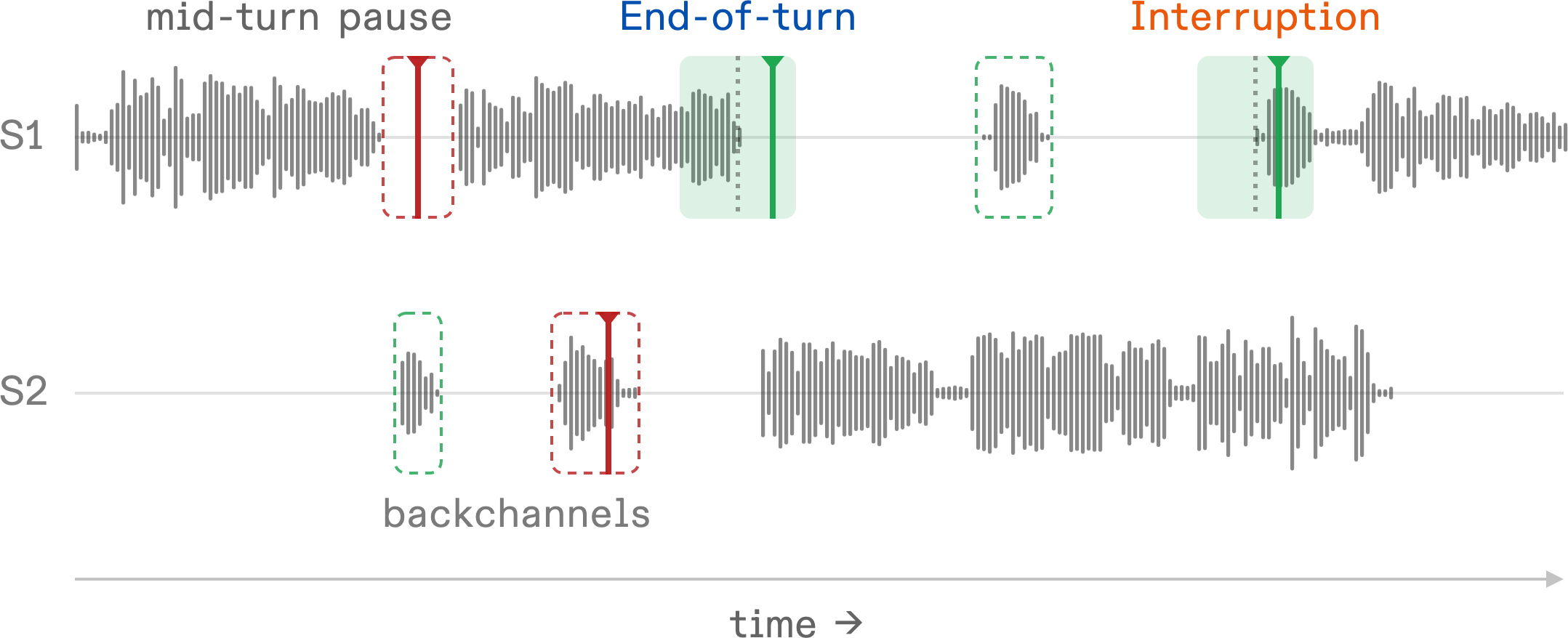}
\caption{The turn-taking events \bench scores, for two speakers (S1, S2).
Dotted lines are human-labeled gold events. Solid lines are model-emitted events, which are true positives (green)
inside the solid scoring windows and false positives (red) inside the dashed
scoring windows. There are two successful detections, as well as two false
positives during a \emph{mid-turn pause} and on a \emph{backchannel}.}
\label{fig:phenomena}
\end{figure}

A similar issue affects evaluation. Full-duplex benchmarks exist \cite{fdbv1, fdbv2, fdbv3}, but each defines backchannels, fillers, and interruptions idiosyncratically, and the definitions are not rooted in linguistic conversation analysis. Comparing systems across benchmarks is therefore difficult \cite{arora2025talking, inoue2024multilingual}, as reported gains may be only an artifact of the annotation convention.

We introduce \bench\footnote{Public leaderboard, dev-set scoring, and interactive conversation viewer: \url{https://turnbench.sesame.com}\label{fn:site}} to address both issues. The benchmark contains dyadic conversations balanced over six conversation types, labeled with event categories grounded in conversation analysis \cite{sacks1974simplest, duncan1972, yngve1970}, and a single protocol that can score disparate system implementations across the same annotations. Our contributions are:
\begin{itemize}
    \item A 30-hour corpus of dual-channel dyadic speech (154 dialogues, 106 voice actors in 53 pairs), balanced across six interaction types, triple-annotated with a fine-grained, analytically grounded taxonomy at strong agreement (Fleiss $\kappa = 0.78$), and released with speaker metadata.
    \item A 104-hour hand-labeled training set, annotated under the same protocol.
    \item A web platform\textsuperscript{\ref{fn:site}} with a public leaderboard, self-serve scoring on the dev split, and a conversation viewer allowing inspection of both labels and predictions with audio playback.
    \item A unified protocol for end-of-turn and interruption detection. We use it to benchmark 14 systems, including rule-based detectors, prompted full-duplex models, semantic and codec endpointers~\cite{UdupaWSC25}, voice activity projection~\cite{ekstedt2022vap, inoue2024multilingual}, and a supervised predictor~\cite{arora2025talking}.
    \item A per-type analysis of 14 open- and closed-source systems. End-of-turn recall stays roughly constant over conversation types, while interruption false positives are type-dependent and most frequent in casual conversation, where backchannels are common. The strongest system, VAP, reaches 0.845 end-of-turn recall and 0.945 interruption recall at median latencies of 368\,ms and 994\,ms.
\end{itemize}

%% file: relwork.tex
\section{Related Work}

\label{sec:related}

Our annotation taxonomy is grounded in conversation analysis, which identifies \emph{Transition-Relevance Places} (TRPs), the structural points at which speaker change is legitimate \cite{sacks1974simplest, duncan1972}, and \emph{backchannels}, listener vocalizations that signal attention without claiming the floor \cite{yngve1970, ward2022interaction}. We adopt both definitions.

\begin{table}[!h]
\centering
\small
\caption{Comparison of dialogue corpora and turn-taking benchmarks.
  DA = dialogue acts, BC = backchannels,
  EOT = end-of-turn, FT = floor transitions, INT = interruptions,
  Disfl = disfluency. \textit{Reg.} = number of registers (conversation types).
  \textit{Ling.} = linguistically grounded annotation scheme.
  \textit{Eval.} = includes model evaluation (O = open-source, C = closed-source).}
\label{tab:corpora-comparison}
\resizebox{\columnwidth}{!}{%
\begin{tabular}{llcccc}
\toprule
\textbf{Resource} & \textbf{Annotations} &
\textbf{Hours} & \textbf{Reg.} &
\textbf{Ling.} & \textbf{Eval.} \\
\midrule
Switchboard \cite{godfrey1992switchboard}
  & Turns, DA, Disfl, prosody     & $\sim$260  & 1 & \xmark & \xmark \\
Switchboard NXT \cite{calhoun2010nxt}
  & + syntax, coref, speech acts  & $\sim$260  & 1 & \cmark & \xmark \\
Fisher \cite{cieri2004fisher}
  & Transcripts                   & $\sim$1960 & 1 & \xmark & \xmark \\
Map Task \cite{anderson1991maptask}
  & DA, turns                     & $\sim$18   & 1 & \cmark & \xmark \\
AMI \cite{carletta2006ami}
  & DA, turns, summaries          & $\sim$100  & 1 & \cmark & \xmark \\
ICSI \cite{janin2003icsi}
  & DA (MRDA)                     & $\sim$72   & 1 & \cmark & \xmark \\
Santa Barbara \cite{dubois2000santabarbara}
  & Transcripts, prosody          & $\sim$20   & 1 & \cmark & \xmark \\
CHiME-5 \cite{barker2018chime5}
  & Transcripts                   & $\sim$50   & 1 & \xmark & \xmark \\
DiPCo \cite{vansegbroeck2020dipco}
  & Transcripts                   & $\sim$5    & 1 & \xmark & \xmark \\
NOTSOFAR-1 \cite{vinnikov2024notsofar}
  & Transcripts, diarization      & $\sim$30   & 1 & \xmark & \xmark \\
DRAL \cite{ward2023dral}
  & Interaction style             & $\sim$2    & 1 & \xmark & \xmark \\
EdAcc \cite{sanabria2023edacc}
  & Transcripts                   & $\sim$40   & 1 & \xmark & \xmark \\
CANDOR \cite{reece2023candor}
  & Transcripts, turn models      & $\sim$850  & 1 & \xmark & \xmark \\
SSSD \cite{sheikh2025sssd}
  & Transcripts                   & $\sim$727  & 1 & \xmark & \xmark \\
CASPER \cite{xiao2025casper}
  & Transcripts                   & $\sim$100  & 1 & \xmark & \xmark \\
otoSpeech \cite{otospeech2025}
  & None                          & 280        & - & \xmark & \xmark \\
\midrule
FDB-v1 \cite{fdbv1}
  & BC, turns, pauses, INT        & $<$1       & 1 & \xmark & O \\
FDB-v2 \cite{fdbv2}
  & Turn fluency, task            & -          & 4 & \xmark & O \\
FDB-v3 \cite{fdbv3}
  & Disfl (5 types), tool use     & -          & 4 & \xmark & O+C \\
Talking Turns \cite{arora2025talking}
  & BC, EOT, turn change          & -$^\dag$   & 1 & \xmark & O+C \\
\midrule
\bench (ours)
  & BC, EOT, FT, INT              & 30         & \textbf{6} & \cmark & O+C \\
\bottomrule
\multicolumn{6}{l}{$^\dag$ Evaluated on a held-out subset of Switchboard.}
\end{tabular}
}
\end{table}

Evaluation frameworks share a related limitation: their categories are not derived from the conversation-analytic definitions motivating the phenomena they target. The Full-Duplex-Bench series \cite{fdbv1, fdbv2, fdbv3} advanced reproducible full-duplex evaluation, but with backchannel, filler, and interruption definitions broader than fine-grained turn-taking requires. Talking Turns \cite{arora2025talking} grounds its evaluation more tightly but covers a single register (telephone), with the models it evaluates performing near chance on backchannels, and none reliably detecting end-of-turn. \bench bases its taxonomy on conversation analysis \cite{sacks1974simplest, ward2022interaction}, covers six conversation types, and enables direct comparison of heterogeneous systems through a unified evaluation protocol (Table~\ref{tab:corpora-comparison}).

%% file: corpus.tex
\section{Corpus Construction}
\label{sec:corpus-construction}

We build \bench around a central hypothesis: conversational \textit{type} (e.g., \ctype{Casual}, \ctype{Argumentative}) is a manipulable variable, set at recording time, that elicits measurably different behavior from humans (\S\ref{sec:corpus-analysis}) and from systems (\S\ref{sec:results}). This motivates the per-type decomposition in our evaluation (\S\ref{sec:evaluation}).

\subsection{Conversation Types}
\label{sec:conversation-types}
Each recording session is assigned one of six conversation types describing its dominant interaction style, following interaction-style literature \cite{ward2023dral, ward2022interaction}.
\begin{itemize}
    \item \ctype{Casual}: unstructured social talk, with topic drift, humor, and many backchannels.
    \item \ctype{Task-Oriented}: goal-directed exchange with clarifications.
    \item \ctype{Instructional}: an expert guides a learner (asymmetric turns, confirmation backchannels).
    \item \ctype{Collaborative}: shared reasoning with no fixed answer (frequent overlap and cooperative interruptions).
    \item \ctype{Argumentative}: structured disagreement (longer turns, fewer backchannels, and competitive interruptions).
    \item \ctype{Narrative}: one speaker tells a story to an active listener (backchannels, little floor competition).
\end{itemize}
The corpus is roughly balanced across types (13--21\% each).

\subsection{Event Taxonomy}
\label{sec:event-taxonomy}

Annotators label time-localized events from a fine-grained inventory of 17 categories grounded in the conversation-analysis literature \cite{sacks1974simplest, duncan1972, yngve1970}. For evaluation, these fine labels are collapsed into seven \emph{canonical} categories (Table~\ref{tab:taxonomy}). Fine annotations are preserved in the released data.

\begin{table}[!tbp]
\centering
\scriptsize
\setlength{\tabcolsep}{3pt}
\renewcommand{\arraystretch}{1.1}
\caption{Canonical event taxonomy and its mapping from the 17 fine annotator labels.}
\label{tab:taxonomy}
\begin{tabular}{@{}>{\raggedright\arraybackslash}p{2cm}>{\raggedright\arraybackslash}p{\dimexpr(\columnwidth-2cm-12pt)*49/100\relax}>{\raggedright\arraybackslash}p{\dimexpr(\columnwidth-2cm-12pt)*51/100\relax}@{}}
\toprule
\textbf{Canonical} & \textbf{Definition} & \textbf{Source fine labels} \\
\midrule
\evt{Turn} & Takes or holds the floor, including overlapping floor-holding speech. & Normal Turn $\cdot$ Strong Floor Hold $\cdot$ Bounded Response $\cdot$ Filler $\cdot$ Overlap \\
\addlinespace[3pt]
\evt{Interruption} & Listener vocalization that takes the floor from the current speaker mid-turn. & Floor-Taking Competitive Interruption $\cdot$ Floor-Taking Cooperative Interruption \\
\addlinespace[3pt]
\evt{Non-floor-taking Interruption} & Interruption attempt that does not take the floor, indistinguishable from \evt{Interruption} at onset. & Non-Floor-Taking Competitive Interruption $\cdot$ Non-Floor-Taking Cooperative Interruption \\
\addlinespace[3pt]
\evt{Backchannel} & Listener vocalization that does not claim the floor, regardless of duration. & Acknowledgement Backchannel $\cdot$ Continuer Backchannel $\cdot$ Reaction Backchannel \\
\addlinespace[3pt]
\evt{Laughter} & Laughter. & Laughter \\
\addlinespace[3pt]
\evt{Awkward Silence} & Marked silence within the interaction not in a turn. & Awkward Silence \\
\addlinespace[3pt]
\evt{NonContent} & Non-speech or non-linguistic audio. & Non-Speech Noise $\cdot$ Channel Bleed $\cdot$ Speech, Non-Linguistic \\
\bottomrule
\end{tabular}
\end{table}

\subsection{Recording Protocol}
\label{sec:recording}

All sessions were recorded in a professional studio with speakers in separate sound-isolated booths using multi-pattern condenser microphones at 48\,kHz/32-bit. A director and sound engineer were present throughout. Any recordings below a minimum quality threshold were discarded before annotation, and released audio is in FLAC, per-speaker.

\subsection{Sessions and Participants}
\label{sec:participants}

The corpus contains approximately 30 hours of dual-channel speech across 154 dialogues (mean duration ${\sim}11.7$ min), recorded by 106 voice actors in 53 pairs, pre-acquainted where possible. Each session begins with a 2--3 minute unrecorded warm-up where the conversation type and a starting topic are set, including participant roles when needed (e.g., explainer and learner in \ctype{Instructional}). The instructions never mention either fine-grained turn-taking behavior, such as interruptions, backchannels, and overlap, or broader personality directions like being dominant or yielding. No further instructions are given once recording starts, and topics are free to drift. The release includes each speaker's gender and an anonymized identifier.

\subsection{Annotation Procedure}
\label{sec:annotation-procedure}

Audio is segmented by Voice Activity Detection (VAD) into candidate segments, with automatic transcripts provided as a reference (annotators trust the audio over the transcript on conflict). Each segment is reviewed in context and tagged with a single fine label (Table~\ref{tab:taxonomy}) by three independent annotators, in a single pass with no relabeling of prior sessions to avoid calibration drift.

Annotators need not have formal linguistics training, with native or near-native comprehension qualifying. Each self-reports accent and regional background at intake for disagreement analysis. Before annotating, each annotator completes a short training protocol where they are given the guidelines, audio examples emphasizing the interruption vs.\ smooth-transition distinction, and a qualification test against reference labels.

\section{Corpus Analysis}
\label{sec:corpus-analysis}

\begin{table*}[!tbp]
\centering
\footnotesize
\setlength{\tabcolsep}{2.2pt}
\renewcommand{\arraystretch}{1.25}
\caption{Per-conversation-type overview. The dynamics columns are computed on the raw annotations: turn length is the mean turn duration (s), and speaker changes and overlap duration (s) are per dialogue. The \emph{Consensus Gold} columns describe Kept (\%) = the share of annotator events matching a same-label consensus event within $\pm$200\,ms at both endpoints (\S\ref{sec:iaa}); Dev/Test = speaker-disjoint dialogue counts; $N$ = total consensus events across all canonical labels; TURN/INT/BC = counts of retained \evt{Turn}, \evt{Interruption}, \evt{Backchannel} events. Switchboard~\cite{godfrey1992switchboard} is a single-register, single-annotator reference.}
\label{tab:type-overview}
\begin{tabular}{l ccccccc @{\hskip 9pt} ccccccc}
\toprule
& \multicolumn{7}{c}{\textbf{Raw Per-Annotator Dynamics}} & \multicolumn{7}{c}{\textbf{Consensus Gold}} \\
\cmidrule(lr){2-8}\cmidrule(lr){9-15}
Conversation Type & Events/min & Words/min & Turn Length & Speaker Changes & BC/min & INT/min & Overlap & Dev & Test & Kept (\%) & $N$ & TURN & INT & BC \\
\midrule
\ctype{Argumentative} & 17.0 & 206 & 10.0 & 30.7 & 3.09 & 2.48 & 20.2 & 8 & 24 & 83.5 & 5701 & 2062 & 377 & 1130 \\
\ctype{Casual}        & 20.9 & 211 &  8.1 & 44.9 & 5.20 & 2.05 & 29.7 & 7 & 22 & 83.6 & 7044 & 2644 & 171 & 1882 \\
\ctype{Collaborative} & 21.4 & 207 &  7.9 & 42.6 & 3.97 & 2.64 & 38.2 & 6 & 20 & 84.9 & 6185 & 2202 & 206 & 1208 \\
\ctype{Instructional} & 17.6 & 202 & 10.5 & 31.3 & 4.30 & 1.52 & 12.9 & 6 & 19 & 89.1 & 4898 & 1698 & 147 & 1240 \\
\ctype{Narrative}     & 18.8 & 198 & 10.2 & 32.5 & 4.76 & 1.50 & 23.0 & 5 & 15 & 87.8 & 4298 & 1496 & 85 & 1134 \\
\ctype{Task-Oriented} & 18.3 & 205 &  8.9 & 34.1 & 5.00 & 1.75 & 16.7 & 6 & 16 & 87.8 & 4694 & 1689 & 165 & 1323 \\
\midrule
All                   & 19.0 & 205 & 9.2 & 36.2 & 4.32 & 2.04 & 23.7 & 38 & 116 & 85.8 & 32820 & 11791 & 1151 & 7917 \\
\textit{Switchboard}~\cite{godfrey1992switchboard} & 18.5 & 198 & 9.4 & 49.3 & 3.90 & 0.86 & 9.4 & -- & -- & -- & -- & -- & -- & -- \\
\bottomrule
\end{tabular}
\end{table*}

\subsection{Consensus and Inter-Annotator Agreement}
\label{sec:iaa}

We combine the three annotations per segment into a single gold by majority consensus when at least two annotators give a segment the same canonical label and their endpoints agree within $\pm$200\,ms. Each endpoint of the gold span is the median across the agreeing annotators, and spans with no majority become \emph{excluded intervals}, dropped by the scorer. The two evaluation tracks use this gold differently. EOT scoring need only know if a span belongs to the current speaker's turn (the \emph{turn view}), while INT scoring keeps the full canonical label (the \emph{label view}). We release the raw annotations and rebuild the gold deterministically at scoring time.

Agreement is high. At the frame level (100\,ms) after the canonical mapping, pairwise Cohen's $\kappa$ is 0.77--0.80 and Fleiss' $\kappa$ is 0.78. Event onsets agree at a boundary F1 of 0.94--0.96 within $\pm$200\,ms, and 85.8\% of annotator events survive gold filtering across the corpus (Table~\ref{tab:type-overview}). Events that do not pass mostly agree on timing but annotators disagree over the label itself. Typically, this is whether an overlapping vocalization represents a shift in the conversational floor (i.e., differentiating between turn, interruption, and backchannel).\footnote{The canonical mapping merges interruption stance (cooperative vs.\ competitive) and backchannel subtype because no public baseline predicts either one. \evt{Non-floor-taking Interruption} remains a separate category because at onset it cannot be distinguished from a floor-taking interruption, and its label is ultimately dependent on future information revealed seconds later. Because of this nuance, we treat these interruption attempts as excluded intervals instead of negatives (\S\ref{sec:int}).}

From these two views we derive the gold the evaluation uses. On the turn view, an \emph{end-of-turn} is anchored at every turn-segment end where the floor passes to the other speaker, yielding 8{,}197 EOT anchors and 4{,}254 mid-turn pause negatives across the corpus. On the label view, consensus floor-taking interruption onsets yield 1{,}151 INT anchors, scored against 17{,}511 \evt{Backchannel} and \evt{NonContent} negatives.

\subsection{Dataset Statistics}
\label{sec:stats}

Table~\ref{tab:type-overview} summarizes the turn-taking dynamics per conversation type alongside consensus retention and the dev/test split. Retention is lower in the higher-overlap types. \ctype{Instructional}, \ctype{Narrative}, and \ctype{Task-Oriented} retain about 88--89\% of annotator events, versus \ctype{Casual}, \ctype{Collaborative}, and \ctype{Argumentative}, which retain about 84--85\%.

The six types differ substantially in turn-taking dynamics (Table~\ref{tab:type-overview}), as predicted in \S\ref{sec:conversation-types}: \ctype{Argumentative} dialogues contribute the most interruption events in the consensus gold (377), \ctype{Casual} and \ctype{Collaborative} show the fastest exchange and highest overlap, while \ctype{Instructional} and \ctype{Narrative} have the longest turns and fewest interruptions. These patterns motivate the per-type breakdowns in \S\ref{sec:evaluation}.

Over the entire corpus on the consensus view (\S\ref{sec:iaa}), humans begin turn transfers \emph{before} the current turn ends, with a median offset of $-281$\,ms, or $-151$\,ms when floor-taking interruptions are excluded. After a floor-taking interruption begins, the interrupted speaker keeps speaking for a median 1.48\,s before ceding the floor.

As a single-register spontaneous reference, Table~\ref{tab:type-overview} includes Switchboard-1~\cite{godfrey1992switchboard}, its events derived from word alignments rather than hand-annotated. Comparing \bench{} vs.\ Switchboard throughout, both aggregate tempo and speaker timing distributions are close (events/min 19.0 vs.\ 18.5 and WPM 205 vs.\ 198). Among transfers that do not start in overlap, inter-speaker gap medians are 0.38\,s vs.\ 0.36\,s (KS $D{=}0.05$), and intra-speaker pause medians are both 0.51\,s (KS $D{=}0.16$). \bench transfers the floor in overlap more often (64\% vs.\ 43\%), as expected from its interruption-dense conversation types.

\subsection{Train / Dev / Test Splits}
\label{sec:splits}

\bench is released in three parts. A training set provides $\sim$104\,h of hand-labeled full-duplex dialogue, speaker-disjoint from the benchmark. It is annotated under the same protocol described in \S\ref{sec:annotation-procedure}\footnote{\url{https://huggingface.co/datasets/otoearth/otoSpeech-full-duplex-turn-104h}} and used for any baselines reported under the \bench-trained condition (\S\ref{sec:baselines}). The \bench corpus itself is partitioned by a 25/75 speaker-disjoint, type-balanced dev/test split (Table~\ref{tab:type-overview}). The dev split (38 dialogues) is released publicly with labels\footnote{\url{https://huggingface.co/datasets/mundo-ai/turn-benchmark-dev}}, while the test split (116 dialogues) is released without labels, audio-only.\footnote{\url{https://huggingface.co/datasets/mundo-ai/turn-benchmark-test}} We additionally release the per-annotator tracks and annotator metadata, so disagreement can be used as signal rather than discarded. All data is distributed under a custom non-commercial license prohibiting voice cloning\footnote{\url{https://huggingface.co/datasets/mundo-ai/turn-benchmark-dev/blob/main/LICENSE}} (\S\ref{sec:ethics}).

%% file: eval.tex
\section{The \bench Benchmark}
\label{sec:evaluation}

\bench defines two evaluation tracks, \textbf{End-of-Turn} (EOT, \S\ref{sec:eot}) and \textbf{Interruption} (INT, \S\ref{sec:int}). Both are boundary-detection tasks scored per speaker and per conversation (Fig.~\ref{fig:phenomena}). A submission is a JSON file that lists, per conversation and per speaker, the timestamps at which each event occurs.\footnote{Full schema and format: \url{https://turnbench.sesame.com/dev}.} Each timestamp is a causal commit time determined by the model. The scorer does not sweep a threshold, so every submission carries its own operating point. The submission format, gold construction, and scoring code are contained in the paper's GitHub repository\footnote{\url{https://github.com/SesameAILabs/turnbench}}.

\subsection{End-of-Turn (EOT)}
\label{sec:eot}
\paragraph{Task} An EOT is the time at which the conversational floor leaves a speaker (the arrival of a TRP, \S\ref{sec:related}). Per speaker, a submission lists EOT times, which the scorer matches against gold turn segment-ends from the consensus \emph{turn view} (\S\ref{sec:iaa}). Because annotation is VAD-segmented, a turn is split into many short \evt{Turn} segments, so most segment-ends are mid-turn pauses rather than real EOTs.

\paragraph{Positives and negatives} A segment-end is a positive EOT if the floor passes to the other speaker (or it is the speaker's last turn), anchored at the segment end, and a negative EOT (mid-turn pause) if the same speaker resumes with no handover. A negative EOT is a span from the segment end, truncated at the first contrary evidence (an excluded interval, the speaker's own non-floor vocalization such as a backchannel, or an other-speaker interruption); other-speaker backchannels do not truncate, since they signal the speaker still holds the floor. A negative EOT span never extends past the speaker's resumption, so a real EOT cannot fall inside one and a correct model is not penalized.

\subsection{Interruption (INT)}
\label{sec:int}

\paragraph{Task} An interruption is a floor entry by the listener away from a TRP, taking the floor from the current speaker mid-turn (a barge-in). Per speaker, a submission lists interruption times on the interrupter's channel, and the gold anchor is the interrupter's onset.

\paragraph{Positives and negatives} A positive INT is a consensus floor-taking \evt{Interruption} onset (2-of-3, on the interrupter's channel). A negative INT span is the time extent of a \evt{Backchannel} or \evt{NonContent} event, neither of which causes a floor transition. Other listener events (\evt{Laughter}) are neither positives nor negatives.

\paragraph{Excluded} Consensus \evt{Non-floor-taking Interruption} events and floor-taking \evt{Interruption} events without consensus are excluded intervals, not negatives. At onset, a non-floor-taking attempt is indistinguishable from a real interruption, so firing on one is neither rewarded nor penalized.

\subsection{Evaluation Protocol}
\label{sec:protocol}

\paragraph{Submission and scoring} For each gold positive at time $t$, the scorer searches the submission for a matching event in the window $[t-\tau_\text{pre},\,t+\tau_\text{max}]$ ($\tau_\text{pre}=0.25$\,s, $\tau_\text{max}=3.0$\,s), where $\tau_\text{max}$ is the latency deadline and $\tau_\text{pre}$ a matching tolerance, and the earliest unclaimed prediction in the window is a true positive. Inside a negative span, firing counts as at most one false positive and not firing counts as one true negative. Predictions in excluded intervals, as well as outside positive and negative spans, are ignored. Thus, FPR measures firing on scored negative spans.

\paragraph{Metrics} Per task we report recall ($TP/(TP{+}FN)$), false positive rate ($FP/(FP{+}TN)$), and signed latency ($\Delta t = t_\text{pred}-t_\text{gold}$) over matched true positives, at the 10th, 50th, and 90th percentiles. A negative latency means the model committed before the gold boundary. The leaderboard ranks submissions by test recall, subject to a 0.15 false positive rate (FPR) ceiling, looser than the 0.1 dev budget (\S\ref{sec:results}) to account for dev-test generalization. Submissions over the ceiling rank below all qualifiers.

%% file: baselines.tex
\section{Baselines}
\label{sec:baselines}
We benchmark 14 turn-taking systems spanning rule-based, academic, and deployed commercial approaches, over single- and dual-channel inputs. Predictors with an accessible training pipeline are fine-tuned on the \bench training split (pooled with Switchboard for the causal WavLM variants). Published models and commercial tools are scored as-is.

\begin{table*}[!tbp]
\centering
\scriptsize
\setlength{\tabcolsep}{3pt}
\renewcommand{\arraystretch}{1.1}
\caption{Test-set results per baseline and conversation type. EOT and INT (Interruption) sub-columns report recall\,/\,fpr (leading zeros omitted). \textsc{Overall}: pooled test recall\,/\,fpr and median latency (ms), $\Delta t = t_\text{pred}-t_\text{gold}$, per track. --- marks an unsupported track. Bold: best recall per column among models within the 0.1 dev false-positive budget.}
\label{tab:by-type}
\begin{tabular*}{\textwidth}{@{\extracolsep{\fill}}l|cc|cc|cc|cc|cc|cc|cc|cc@{}}
\toprule
& \multicolumn{2}{c|}{\ctype{Argumentative}} & \multicolumn{2}{c|}{\ctype{Casual}} & \multicolumn{2}{c|}{\ctype{Collaborative}} & \multicolumn{2}{c|}{\ctype{Instructional}} & \multicolumn{2}{c|}{\ctype{Narrative}} & \multicolumn{2}{c|}{\ctype{Task-Oriented}} & \multicolumn{2}{c|}{Overall} & \multicolumn{2}{c}{Overall $\Delta t$} \\
\cmidrule(lr){2-3}\cmidrule(lr){4-5}\cmidrule(lr){6-7}\cmidrule(lr){8-9}\cmidrule(lr){10-11}\cmidrule(lr){12-13}\cmidrule(lr){14-15}\cmidrule(lr){16-17}
Baseline & EOT & INT & EOT & INT & EOT & INT & EOT & INT & EOT & INT & EOT & INT & EOT & INT & EOT & INT \\
\midrule
RMS VAD                        & .71/.64 & 1.0/.39 & .78/.70 & 1.0/.52 & .66/.58 & 1.0/.38 & .64/.51 & .99/.42 & .75/.70 & .97/.41 & .79/.66 & 1.0/.55 & .718/.632 & .996/.445 & $-$117 & 123 \\
\midrule
OpenAI Realtime (Server VAD)   & .96/.47 & .99/.38 & .94/.50 & .98/.54 & .95/.60 & .99/.39 & .96/.50 & .99/.43 & .96/.57 & 1.0/.45 & .95/.51 & 1.0/.54 & .955/.525 & .990/.458 & 282 & 184 \\
OpenAI Realtime (Semantic VAD) & .27/.02 & .50/.24 & .29/.01 & .37/.29 & .35/.02 & .46/.24 & .30/.03 & .61/.27 & .32/.03 & .53/.28 & .27/.00 & .48/.32 & .303/.018 & .484/.271 & 793 & 196 \\
Kyutai SVAD                    & .79/.08 & .90/.06 & .73/.04 & .90/.12 & .78/.08 & .87/.07 & .79/.06 & .92/.07 & .76/.05 & .91/.08 & .79/.04 & .90/.07 & .773/.059 & .898/.081 & 1007 & 559 \\
SmartTurn v3                   & .76/.05 & .12/.09 & .73/.03 & .04/.11 & .75/.04 & .11/.06 & .78/.05 & .17/.13 & .73/.06 & .12/.08 & .75/.05 & .11/.10 & .752/.047 & .107/.093 & 1017 & 159 \\
\midrule
VAP                            & \textbf{.88/.04} & \textbf{.95/.09} & \textbf{.82/.07} & \textbf{.94/.13} & \textbf{.84/.06} & \textbf{.94/.12} & \textbf{.85/.03} & \textbf{.93/.09} & \textbf{.82/.07} & \textbf{.93/.09} & \textbf{.86/.05} & \textbf{.98/.12} & \textbf{.845/.055} & \textbf{.945/.107} & 368 & 994 \\
Mimi-EP                        & .77/.05 & .89/.08 & .76/.09 & .92/.13 & .81/.09 & .92/.10 & .81/.06 & .91/.09 & .76/.12 & .84/.12 & .78/.07 & .92/.11 & .782/.078 & .899/.106 & 645 & 1007 \\
ESPnet TT-pred.\               & .83/.09 & .60/.08 & .81/.07 & .56/.11 & .82/.09 & .50/.06 & .85/.06 & .58/.07 & .81/.08 & .69/.08 & .85/.06 & .53/.07 & .826/.078 & .573/.080 & 862 & 210 \\
ESPnet TT-pred.\ (per-ch.)     & .70/.08 & .60/.12 & .75/.09 & .71/.19 & .69/.08 & .58/.13 & .68/.07 & .44/.09 & .70/.09 & .70/.11 & .74/.07 & .63/.15 & .711/.081 & .611/.135 & 730 & 859 \\
\midrule
WavLM-Base (causal)            & .36/.06 & .77/.08 & .39/.06 & .80/.17 & .44/.07 & .87/.11 & .42/.06 & .86/.09 & .42/.07 & .78/.10 & .39/.05 & .90/.10 & .403/.061 & .820/.111 & 701 & 580 \\
WavLM-Large (causal)           & .37/.07 & .60/.06 & .37/.04 & .60/.14 & .43/.06 & .67/.10 & .46/.05 & .68/.06 & .43/.04 & .58/.10 & .40/.05 & .75/.09 & .408/.054 & .637/.094 & 683 & 832 \\
WavLM-Large (anchor)           & .81/.07 & .89/.06 & .77/.03 & .86/.09 & .80/.06 & .86/.06 & .83/.07 & .88/.04 & .80/.05 & .85/.04 & .81/.04 & .83/.04 & .800/.054 & .868/.054 & 1076 & 1412 \\
\midrule
Gemini 3.1 Live                & .71/.02 & --- & .62/.02 & --- & .68/.03 & --- & .65/.02 & --- & .61/.03 & --- & .65/.01 & --- & .657/.022 & --- & 1234 & --- \\
Moshi                          & .21/.04 & --- & .24/.03 & --- & .23/.04 & --- & .29/.06 & --- & .20/.07 & --- & .24/.03 & --- & .233/.044 & --- & 702 & --- \\
\bottomrule
\end{tabular*}
\end{table*}


\subsection{Rule-Based Heuristics}
The RMS VAD fires whenever channel energy crosses a fixed threshold. It commits an EOT when the speaker's channel falls silent and an interruption when the listener's channel becomes active during the speaker's turn. This baseline uses no linguistic information and is the floor of the benchmark.

\subsection{Open and Commercial Tools}
OpenAI Realtime's Server VAD and Semantic VAD are the acoustic and semantic endpointing modes, respectively, of a deployed API\footnote{\texttt{gpt-realtime}, evaluated June 2026.}: Server VAD commits on silence duration alone, while Semantic VAD adds a turn-detection model that waits longer when linguistic content suggests the turn is unfinished. We also evaluate Kyutai SVAD\footnote{The VAD head of \texttt{kyutai/stt-1b-en\_fr-candle}.}, which pairs streaming ASR with a semantic EOT head, and SmartTurn v3\footnote{\texttt{pipecat-ai/smart-turn-v3}, v3.1 ONNX weights.}, which emits a per-chunk turn-completion probability. Neither system has an interruption head, so we invert the turn-ending score on the interrupting speaker's channel ($1 - P(\text{turn ending})$) and commit an interruption where it rises above a threshold.

\subsection{Turn-Taking Predictors}
ESPnet's Switchboard model \cite{arora2025talking} emits a 5-class head so we threshold Turn-change for EOT and Interruption for INT. VAP \cite{ekstedt2022vap, inoue2024realtime, inoue2024multilingual} predicts continuous future voice activity per speaker, firing EOT when a speaker's own floor-hold probability drops and an interruption when the interrupting speaker's rises. The Mimi-based endpointer \cite{UdupaWSC25} emits a 5-class endpointer state over Mimi codec tokens \cite{defossez2024moshi}; we score EOT with $1 - P(\textit{user})$ and interruption with $P(\textit{user})$. We additionally train WavLM-Large (anchor), a predictor on a frozen WavLM-Large frontend \cite{chen2022wavlm}, adapted from the ANCHOR speech-quality framework \cite{tao2026anchor}, emitting a 5-class distribution at 25\,Hz over 4\,s current-anchored windows (bidirectional within a window, 0\,ms effective lookahead). We also test WavLM-Base/Large (causal), which use left context only, and a per-channel ESPnet variant that runs on each speaker's channel instead of the mixed mono input.

\subsection{Full-Duplex Models}
To test whether full-duplex models handle turn-taking zero-shot, we
place the model in the conversation. Twice for each conversation, one speaker's channel is streamed into a live session, with the model standing in for the
other party in real time. Its output audio is then recorded and sample-aligned with the input. We evaluate Gemini 3.1 Live\footnote{\texttt{gemini-3.1-flash-live-preview}, evaluated June 2026,
prompted as a conversational agent (configuration in the repository).} and Moshi
\cite{defossez2024moshi}. Neither exposes turn-taking labels, so we read
decisions from the produced audio: an EOT is committed at each model speech
onset detected by pyannote VAD
\cite{bredin2020pyannote}. Latency for full-duplex models thus includes response generation time and (for Gemini) network delay. We score
full-duplex models on EOT only, since the model cannot be guaranteed to be speaking when an INT event occurs.

%% file: results.tex
\section{Results and Discussion}
\label{sec:results}
We report results on the test set (\S\ref{sec:splits}; 116 conversations) for all baselines (\S\ref{sec:baselines}). Because we require discrete, causal event timestamps in a system's submission, any model emitting continuous probabilities faces a trade-off between latency and FPR when selecting its threshold. We thus fix model operating points using a threshold sweep on the dev set, independently for EOT and INT. Each swept model uses a common commit rule (one event per rising edge above $\theta$, with a 2\,s refractory period) to convert its native output to commit times. Thresholds are quantiles of the model's own probability distribution, so the sweep is scale-invariant. We select the highest-recall threshold $\theta$ whose dev FPR stays within the 0.1 budget (Fig.~\ref{fig:threshold-sweep-dev}), freeze it, and use this operating point to evaluate on the test set.

\noindent\begin{minipage}{\columnwidth}
\centering
\includegraphics[width=\columnwidth]{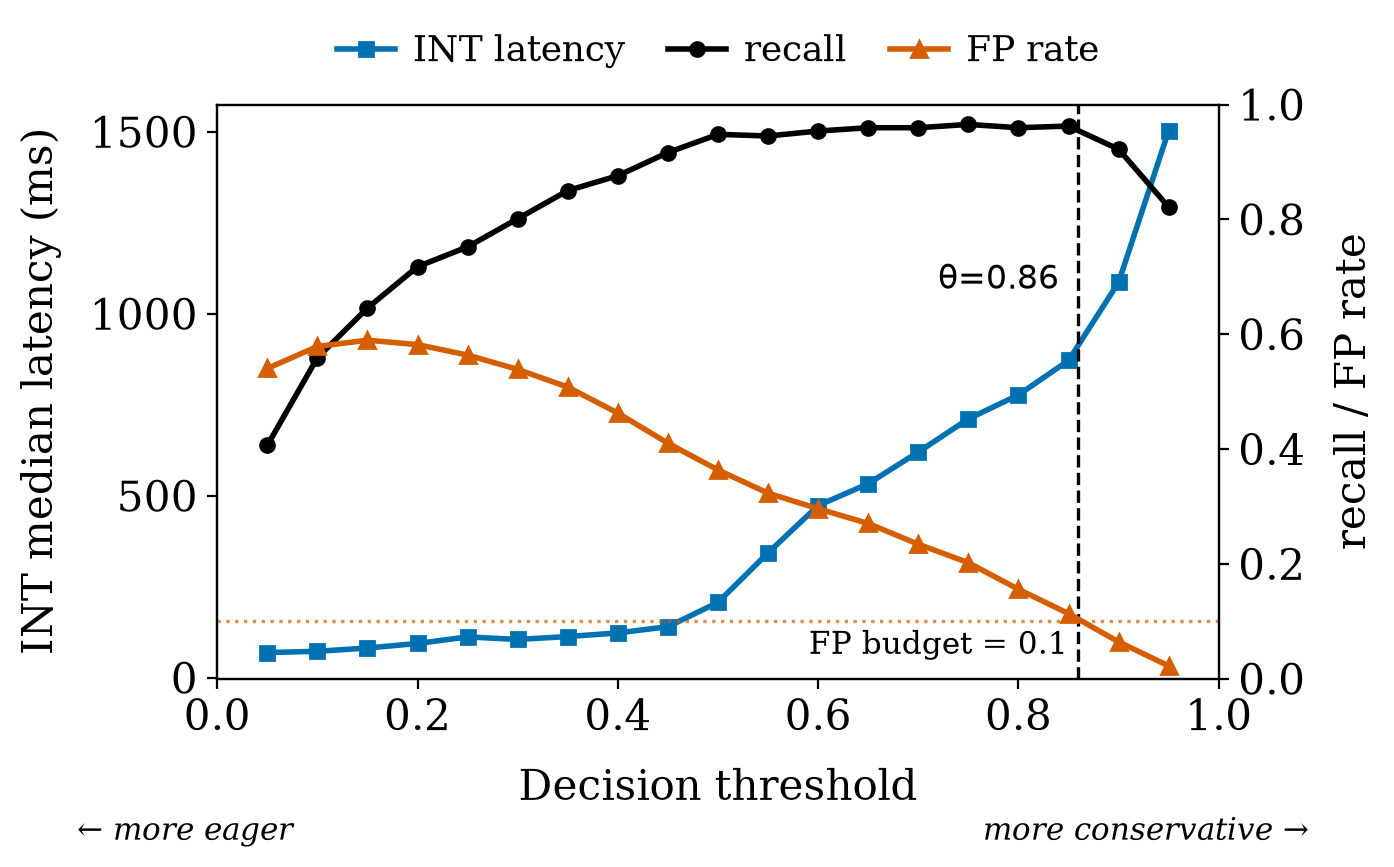}
\captionof{figure}{Interruption threshold sweep on \bench dev for VAP. Median latency (blue, left axis), FPR (orange, right axis), and recall (black) vs.\ decision threshold $\theta$. The dashed line at $\theta{=}0.86$ is the operating point with the highest recall within the 0.1 FPR budget (dotted orange line).}
\label{fig:threshold-sweep-dev}
\end{minipage}

Table~\ref{tab:by-type} reports EOT and INT results across the evaluated baselines and conversation types. Casual INT FPR exceeds Argumentative INT FPR for every model. The RMS VAD yields anticipatory EOT latency ($-117$\,ms) at very high FPRs of 0.51--0.70 (EOT) and 0.38--0.55 (INT) across types, firing on nearly every silence and onset. Among in-budget models, VAP achieves the strongest operating point on both tracks. On EOT: 0.845 recall at 0.055 FPR with 368\,ms median latency. On INT: 0.945 recall at 0.107 FPR with 994\,ms median latency. Gemini sits at the conservative end, with the lowest FPRs (0.01--0.03), 0.61--0.71 recall, and the highest EOT latency (1234\,ms). Moshi stays within FPR budget but at low 0.20--0.29 EOT recall, progressively falling silent over long sessions. No in-budget system approaches the human reference (\S\ref{sec:stats}) of beginning transfers a median 151\,ms before the turn ends.

\paragraph{Acoustic versus linguistic endpointing} Turn-taking theory predicts acoustic-only detectors should be overeager, as they do not account for linguistic cues. RMS VAD and OpenAI Server VAD saturate recall at FPRs above budget, while the linguistically informed endpointers, projection models, and supervised predictors stay in budget. The two OpenAI modes differ only in endpointing yet differ sharply in FPR.

\paragraph{Interruption detection trades speed for selectivity} Systems that commit INT quickly focus on the interrupter's speech onset: SmartTurn v3 (159\,ms), ESPnet (210\,ms), and both OpenAI modes. At its onset an interruption is indistinguishable from a backchannel, so speed costs either FPR or recall. OpenAI Server VAD fires on nearly half the backchannels (0.458 FPR) while SmartTurn recovers only 0.107 of interruptions. VAP (994\,ms), Mimi-EP (1007\,ms), and WavLM-Large (anchor) show the inverse. They wait past onset to discriminate, reaching 0.87--0.95 recall at 0.05--0.11 FPR. Kyutai SVAD sits between (559\,ms, 0.898 recall, 0.081 FPR). Onset-driven systems commit INT faster than their own EOT, while VAP, Mimi-EP, and WavLM-Large show the opposite.

%% file: conclusion.tex
\newpage
\section{Conclusion}
\looseness=-1
We present \bench, a multi-domain turn-taking benchmark grounded in conversation analysis, with 30 hours of triple-annotated dyadic speech, a 104-hour training set, and a reproducible protocol for end-of-turn and interruption detection. Across 14 heterogeneous turn-taking systems, we find end-of-turn recall is largely stable across conversation types, while interruption false positives are not. No system is simultaneously fast, selective, and high-recall.

\looseness=-1
\bench has limitations. All dialogues are English, studio-recorded, and dyadic. Majority consensus drops events annotators disagree on, though that disagreement itself carries signal. We also lack a method for scoring interruption in full-duplex models. We leave multilingual, non-studio recordings, analysis of the fine-grained or per-annotator labels, and full-duplex interruption evaluation to future work.

%% file: ethics.tex
\section{Ethics Statement}
\label{sec:ethics}

All voice actors and annotators gave written informed consent for recording, annotation, and public release, were paid market rates, and could withdraw before release. Released data includes only gender and an anonymized identifier, under a non-commercial license prohibiting voice cloning (\S\ref{sec:splits}).